\documentclass[twocolumn,showpacs,amsmath,amssymb,aps,prl,superscriptaddress]{revtex4}
\usepackage{graphicx}     
\usepackage{bm}

\begin{document}

\title{Subarea law of entanglement in nodal
fermionic systems}

\author{Letian Ding}
\affiliation{Department of Physics and Astronomy, University of
Southern California, Los Angeles, CA 90089}

\author{Noah Bray-Ali}
\affiliation{Department of Physics and Astronomy,
University of Southern California, Los Angeles, CA 90089}

\author{Rong Yu}
\affiliation{Department of Physics and Astronomy, University of
Southern California, Los Angeles, CA 90089} \affiliation{Department
of Physics and Astronomy, University of Tennessee, Knoxville, TN
37996}
\author{Stephan Haas}
\affiliation{Department of Physics and Astronomy, University of
Southern California, Los Angeles, CA 90089}

\date{\today}

\begin{abstract}
We investigate the subarea law scaling behavior of the block entropy
in bipartite fermionic systems which do not have a finite Fermi
surface. It is found that in gapped regimes the leading subarea term
is a negative constant, whereas in critical regimes with point nodes
the leading subarea law is a logarithmic additive term. At the phase
boundary that separates the critical and non-critical regimes, the
subarea scaling shows power-law behavior.
\end{abstract}

 \maketitle

The study of entanglement across quantum phase transitions has
attracted substantial activity in the physics community. This
interest was initially boosted by the discovery of a connection
between the scaling properties of the concurrence, which is a
measure of entanglement, and the quantum phase transition in the
one-dimensional (1D) transverse Ising model\cite{Osterlohetal02,
OsborneN02}. Subsequently,  the scaling properties of the von
Neumann entanglement entropy (EE), which is defined as $S_L = -{\rm
Tr} \rho_{L^d} \log \rho_{L^d}$  of its reduced density matrix
$\rho_{L^d} = Tr_{U\setminus L^d} |\Psi\rangle \langle \Psi |$ and
$L$ is the length of a contiguous system block $L^d$, show a
dramatic difference between 1D critical and non-critical systems: in
critical regimes it exhibits logarithmic scaling $S_L\propto \log
L$, whereas in non-critical regimes it saturates to a constant as $L
\rightarrow \infty$ \cite{Vidaletal03,Verstraeteetal04}. Conformal
field theory (CFT) tells us that the analytical form of the EE in 1D
critical cases is given by $S_L=\frac{c+\bar{c}}{6}\log_2{L}$, where
$c$ and $\bar{c}$ are the holomorphic and antiholomorphic central
charges which are universal factors for a given class of critical
systems\cite{CalabreseC04,Korepin04}. In this sense, if the
correlation length diverges, the EE exhibits logarithmic scaling
with block size, and if the correlation length is finite, we simply
obtain a constant EE.

In higher dimensions, $d>1$, the scaling behavior of the EE is
complicated and far from fully settled. On one hand, in bosonic
systems it has been shown that the area law $S_L \sim L^{d-1}$ hold
for both critical and non-critical cases. On the other hand, in
fermionic systems, the scaling behavior of EE highly depends on the
topology of the Fermi surface: the area law was shown to be violated
in systems possess a
finite Fermi surface~\cite{fermisurface01,fermisurface02}; 
but it still holds in gapped systems and systems having point
nodes~\cite{ourgroup,barthel,cramer01,letian01}. It is remarkable
that in systems where the area law holds, the link between
correlation length and the scaling behavior of the EE is broken: the
leading term of EE fails to tell if the correlation length diverges
in the system. This is not too surprising when the leading term
follows the area law since an area law only reflects short-range
correlations in the system. However, this motivates us to study the
subarea behavior of the EE, where the link between correlation
length and EE may still be active.

The subarea behavior of EE has been investigated in some 2D systems.
Noticeably, it exhibits some very interesting properties. In
non-critical systems, the common expectation for the subleading
behavior is a constant, which in some cases can be used to
characterize topological order\cite{topo01,topo02,topo03,topo04}.
For critical systems in the universality class of z=2 conformal
quantum critical points, there exists a universal logarithmically
divergent correction to the non-universal area law, which is
determined by the shape of the partition and by the central charge.
Very recently, a universal logarithmically divergent correction in
the 2D random transverse Ising model was also observed~\cite{yu}.
Following their approach, one might expect that a general
logarithmic additive correction generally exists in critical
systems.

In this paper, we examine the behavior of the subleading term of the
EE in a 2D quadratic fermionic Hamiltonian. Our results show that
in the non-critical regimes the leading subarea term is a negative
constant, proportional to the square root of the correlation
length. On the other hand, in quantum critical phases which have only
point nodes in their excitation spectrum the leading subarea law is
found to be a logarithmic additive term. It is interesting to note that the
subarea law shows unusual behavior in the vicinity of the phase
boundary separating the critical and noncritical regimes. According to
the careful numerical analysis discussed below,
this unusual behavior follows a nonadditive
power law $L^\alpha$ with $\alpha<1$.

\begin{figure}
\includegraphics[width=3.2in]{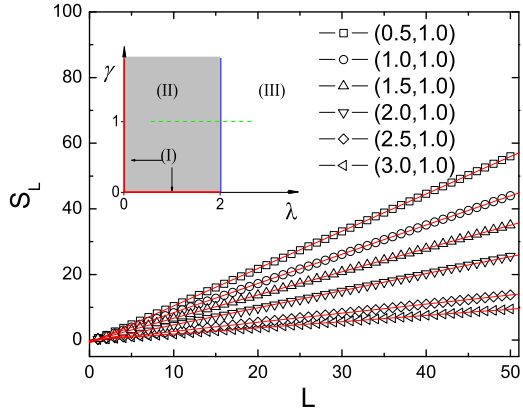}  
\caption{Scaling behavior of the entanglement entropy $S_{L}$ for a
cut through the phase diagram at fixed interaction potential $\gamma
= 1.0$. The chemical potential is varied in the window $\lambda \in
[0.5,3.0]$, thus crossing through phases II and III which do not
have finite Fermi surfaces. The parameters shown in the legend
correspond to $(\lambda,\gamma)$. The inset shows the phase diagram
of the 2D spinless fermion model (Eqn. 1).}
 \label{f0}
\end{figure}

The particular Hamiltonian we consider in this paper is a bilinear spinless
fermionic model on a 2D lattice with pairing interaction
between nearest-neighbor lattice sites\cite{ourgroup},
\begin{eqnarray}
H=\sum_{\langle{\bm i}{\bm j}\rangle} \left[ c^{\dagger}_{{\bm
i}}c_{{\bm j}}-\gamma (c^{\dagger}_{{\bm i}}c^{\dagger}_{{\bm
j}}+c_{{\bm j}}c_{{\bm i}})\right] - \sum_{{\bm i}}2\lambda
c^{\dagger}_{{\bm i}}c_{{\bm i}}. \label{e.Hamilton}
\end{eqnarray}
Here  $\lambda$ is the chemical potential, and $\gamma$ is a p-wave
pairing interaction. The sum $\sum_{\langle {\bm i} {\bm j}\rangle}$
extends over nearest-neighbor pairs. Depending on the parameters
$\gamma$ and $\lambda$, this system has a rich phase diagram,
including metallic, insulating and p-wave superconducting regimes.
In previous work, we distinguished three different phases whose
different signatures are reflected by the scaling properties of the
leading term of the EE. Phase I is the case  $\gamma=0$ and $0\leq
\lambda < 2$. A finite Fermi surface (i.e. line nodes) exists in
this regime.  Phase II is the case with $\{0 < \lambda <
2,\gamma>0\}$. In this regime, only point nodes exist in the
excitation spectrum. Phase III  $\{\lambda
> 2\}$ is an insulating state characterized by a gap in the
excitation spectrum. The resulting phase diagram is shown in the
inset of Fig. \ref{f0}. 
The ground state in this model is known to be a Gaussian state whose
reduced density matrix can be obtained exactly with the usage of
Grassman algebra and a Bogoliubov
transformation\cite{Gaudin60,Peschel03}.  Therefore, one can
calculate the block entropy $S_L$ directly, in the thermodynamic
limit, using the correlation matrix\cite{ourgroup,weifeithesis}.
Previously \cite{ourgroup}, it was found that in phase I the leading
scaling behavior of the EE is $S_L\sim{L\log L}$ and the
coefficients of this term are well described by the analytical form
deduced by the Widom conjecture\cite{fermisurface02}.  In phase III,
the area law $S_L\sim{L}$ holds. Interestingly, in phase II area law
behavior is also observed (Fig. \ref{f0}) although the system has a
diverging correlation length in this case.


To study the subarea law of the EE, we need to first subtract the
dominant area law contribution.
Since we are able to compute the block entropy $S_L$
 directly in the thermodynamic limit, the dominant area
 law contribution depends linearly on the size $L$ of
 the block: $S_L \sim a L,$ for a two-dimensional system.
  This simple linear dependence should be contrasted with
   the dominant, non-monotonic dependence of EE on block size
   in a finite-size system\cite{topo01,topo02,topo03}.
    For example, the dominant contribution to
    the EE of a block of size L in a one-dimensional system with Hamiltonian (\ref{e.Hamilton})
    has the following non-monotic form
    \cite{CalabreseC04}:
\begin{equation}
S_L\sim\frac{c}{3} \log(\frac{N}{\pi}\sin\frac{\pi L}{N}).
\end{equation}
In the thermodynamic limit $N\rightarrow\infty$, this takes the
monotonic form $S_L=\frac{c}{3}\log_2{L}+\ldots.$  Similarly, we
find a monotonic, dominant contribution to the block entropy in two
dimensions, which allows us to extract sub-leading terms reliably.

We will focus on the quantity $S^{Sub}_L=LS_{L+1}-(L+1)S_{L}$, and
we will also examine the second-order derivative
$\frac{d^2S_L}{dL^2}$. It is then straightforward to see that if the
EE follows a behavior $S_L= aL-const+O(\frac{1}{L})$, $S^{Sub}_L\sim
const$ and $\frac{d^2S_L}{dL^2}\sim O(\frac{1}{L^3})$. In contrast,
if the EE scales as $S_L= aL-b\log L+O(1)$, then $S^{Sub}_L\sim
-bL\log(L+1)+b(L+1)\log L\sim b\log L$ and $\frac{d^2S_L}{dL^2}\sim
\frac{b}{L^2}$ for $L \gg 1$. Interestingly, if the EE has a
power-law subleading correction, $S_L=aL-bL^\alpha+O(1)$ with
$\alpha<1$, $S^{Sub}_L\sim bL^\alpha$ and $\frac{d^2S_L}{dL^2}\sim
\frac{b}{L^{2-\alpha}}$.

\begin{figure}
\includegraphics[width=3.6in]{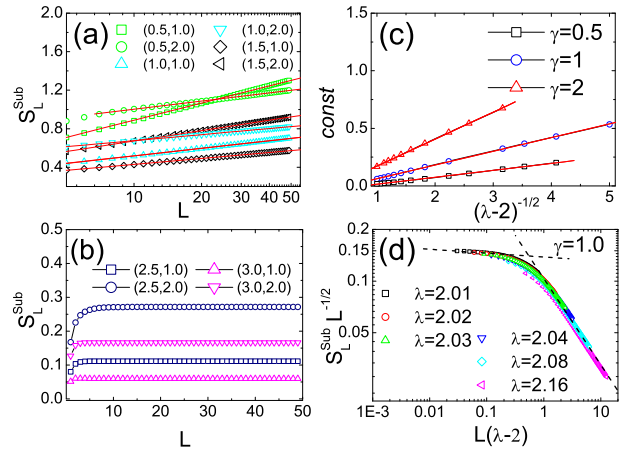}  
\caption{(a) Scaling behavior of  $S^{Sub}_L$  in phase II. (b)
Scaling behavior of $S^{Sub}_L$ in phase III.  (c) The relation
between the constant term $const$ and $\frac{1}{\sqrt{\lambda-2}}$
for fixed finite interaction potential $\gamma = 0.5$,$\gamma =
1.0$,$\gamma = 2.0$ in phase III. (d) Dependence of
$\frac{S^{Sub}_L}{\sqrt{L}}$ v.s. $L(\lambda-2)$ with the fixed
interaction potential $\gamma=1$ in phase III. }
 \label{f1}
\end{figure}

A significant difference of $S^{Sub}_L$ between phase II and phase
III is observed despite the fact that both phases obey an area law
in leading order. The results are shown in Fig. \ref{f1}.  We find
that in phase II $S^{Sub}_L\sim -blogL$, with $b>0$
(Fig.\ref{f1}(a)). In contrast, in phase III, one observes that
$S^{Sub}_L$ quickly converges to a positive constant as  $L$
increases (Fig.\ref{f1}(b)). These results are consistent with the
following scaling forms for the entanglement entropy:
$S_L=aL-blogL+\ldots$, with $b>0$ in phase II, and in phase III
$S_L= aL-const+\ldots$, with $const$ is an positive non-universal
constant which depends on ($\lambda$,$\gamma$) . As noted in Ref.
\cite{fradkin}, the sign of a sub-leading correction to the entropy
may be positive or negative, since the leading term insures that the
entropy is positive, as it should be. In fact, both the conformal
quantum critical points studied in Ref. \cite{fradkin} and the
critical points in Ref. \cite{yu} have subleading logarithmic
corrections with the same sign as found in phase II.

In contrast to the fixed points described in Refs.
\cite{fradkin,yu}, the $z=1$ critical phase II has a non-universal
sub-leading correction\cite{note01}.  For example, in
Fig.\ref{f1}(a), we see that moving around within phase II changes
the slope of $S^{Sub}_L$ dramatically.  Similarly, the sub-leading
correction  is non-universal in phase III, as seen in
Fig.\ref{f1}(b). This is a non-surprising result, considering that
the phase is gapped with a finite correlation length $\xi$.
Interestingly, the absolute value of this correction follows the
same trend as the correlation length $\xi$ in phase III, namely it
increases with increasing pairing interaction $\gamma$ and decreases
if $\lambda$ increases. Therefore, let us now focus on the
relationship between the constant term of $S^{Sub}_L$ in phase III
and the parameters ($\lambda$,$\gamma$). In Fig.\ref{f1}(c), we
examine the relation between the constant term and
$\frac{1}{\sqrt{\lambda-2}}$ at fixed values of the interaction
potential $\gamma = 0.5, 1.0, 2.0$. One finds good linear fits and
observes that the slopes of these lines increase with $\gamma$,
which suggests the following form: $S_L\sim
aL-\frac{b(\gamma)}{\sqrt{\lambda-2}}$ in phase III. Fig.\ref{f1}(d)
gives the relation between $\frac{S^{Sub}_L}{\sqrt{L}}$  and
$L(\lambda-2)$. For fixed $\gamma$, one finds that all curves
collapse. With the the band gap $\Delta_0=\lambda-2$ and the
knowledge that the transition from phase II to phase III has $z=1$
criticality, it is plausible to infer from the above graph that the
subarea law in this regime has a uniform finite-size scaling form:
$S^{Sub}_L\sim L^{\frac{1}{2}}f(\frac{L}{\xi})$, with
$f(x)=x^{-\frac{1}{2}}$ when $x\rightarrow\infty$, and $f(x)=
const.$ when $x\rightarrow0$.


\begin{figure}
\includegraphics[width=3.6in] {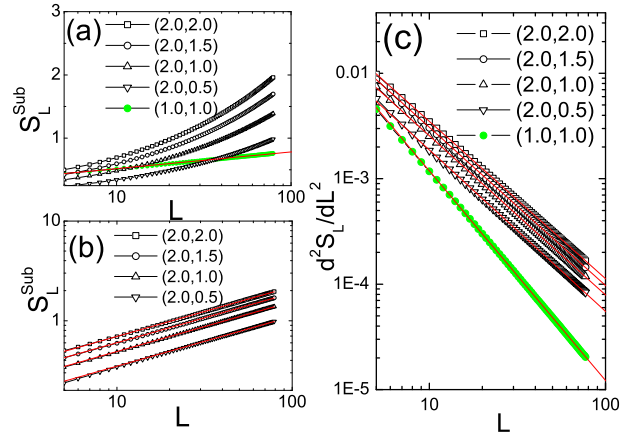}
\caption{(a) Semi-log (Y/logX) graph and (b) Log-log graph of the
scaling behavior of $S^{Sub}_L$ in phase II and its boundary points.
  (c) Scaling behavior of
$\frac{d^2S_L}{dL^2}$ in phase II and its boundary points.}
 \label{f2}
\end{figure}

The critical scaling regime of the boundary between phase II and
phase III provides a striking, {\it universal} correction to the
area law (Fig.\ref{f2}). Here the dominant subleading behavior is
{\it not} logarithmic (See Fig.\ref{f2}(a)).  We find that
$S^{Sub}_L$ exhibits significant curvature as a function of $\log L$
in phase III, in contrast to the linear behavior in phase II
(denoted by filled circles in Fig.\ref{f2}(a)). Furthermore, using
the log-log graph (Fig.\ref{f2}(b)), one finds that $S^{Sub}_L$
follows a power-law relation over almost two decades of scaling.
Finally, the exponent we extract from fitting is close to
$\alpha=0.5$ (eg: $\alpha=0.493\pm0.001$ for $\lambda=2.0$
,$\gamma=2.0$). In fact, all the curves can be fit with the
log-linear law $\log y= .5 logx + b(\gamma)$, with a slope that is
independent of $\gamma$, as long as one remains on the critical line
separating phase II and phase III.  This insensitivity of the
power-law exponent is consistent with universal behavior of this
sub-leading contribution to the entropy.  By dimensional analysis,
 the coefficient $b$ depends on the lattice spacing, which
 is consistent with the observed $\gamma$ dependence of $b$
 in our numerical
 results. We believe this is the first example of a
power-law correction to an area law observed in any dimension.

As we mentioned above, the second-order differentiation
$\frac{d^2S_L}{dL^2}$ of $S_L$  can also be used to detect the
scaling behavior of the subarea law contributions to $S_L$. This
analysis is performed in Fig. \ref{f2}(c)). One finds that
$\frac{d^2S_L}{dL^2}$ follows a perfect inverse power law relation.
However, there exist obvious differences between phase II  and the
boundary region, where the exponent of this inverse power law is not
the same. For phase II, after fitting the curves in a log-log graph,
we obtain $\beta$ very close to $-2.0$.  A typical example with
$\lambda=1.0$, $\gamma=1.0$ is shown in Fig. \ref{f2}(c), the simple
linear fitting gives us $\beta=1.988\pm0.001$ in this case. This
result suggests $S_L\sim aL-b\log L$ and is consistent with the
results from the above analysis of $S^{Sub}_L$ in Fig. \ref{f1}(a).
In the boundary regime, one finds the power factor of
$\frac{d^2S_L}{dL^2}$ to be very close to $-1.5$(eg:
$\beta=-1.509\pm0.002$ for $\lambda=2.0$ ,$\gamma=2.0$), consistent
with the above observation $S^{Sub}_L\sim L^{-0.5}$ for all the four
cases we studied. Therefore, both two numerical methods suggest
unusual power law corrections in the subarea term of the EE.

It is worthy to consider alternative possibilities for the subarea
law in the boundary region apart from power-law corrections. One
possibility is a $\log\log L$ correction as a subleading term. In
this case, $\frac{d^2S_L}{dL^2}$ would decay more rapidly than
$L^{-2}$, and so it cannot explain the results in Fig. \ref{f2}(c)
since in this graph $\frac{d^2S_L}{dL^2}$ is well described by
$L^{-2+\alpha}$ ($\alpha\sim 0.5$) in the boundary regime. A second
possibility would be a term scaling as $-\log^2L$. In this
situation,  it is true that $\frac{-d^2\log^2L}{dL^2}\sim \frac{\log
L}{L^2}-\frac{1}{L^2}$ can give us illusion of power law relation
$L^{-2+\alpha}$ when $L$ is not too large. However, one finds that
$S^{Sub}_L$ does not show a good linear relation on a log-log scale,
and therefore is not consistent with the results in Fig.
\ref{f2}(b). In brief, the unusual non-additive $L^\alpha$ type
subarea law is the most plausible leading candidate according to the
above analysis. Of course, the most straightforward evidence comes
from our observation  $S^{Sub}_L\sim
L^{\frac{1}{2}}f(\frac{L}{\xi})$ ($f(x)=const.$ at $x\rightarrow0$)
in the gapped phase, as shown in Fig.\ref{f1}(d). If one approaches
the phase boundary, $\xi\rightarrow\infty$, the form of subarea law
should be $S^{Sub}_L\sim L^{\frac{1}{2}}$ in analogy to the 1D case
where the finite size scaling form $S_L\sim\log\xi$ in the
non-critical phase changes to $S_L\sim\log{L}$ in the critical phase
in the limit $L>>1$. It is noted that one cannot exclude the
possibility that $\log\log L$ , $\log^2L$, even $\log L$ as the
second leading order subarea-law term.

\begin{figure}
\includegraphics[width=3.6in]{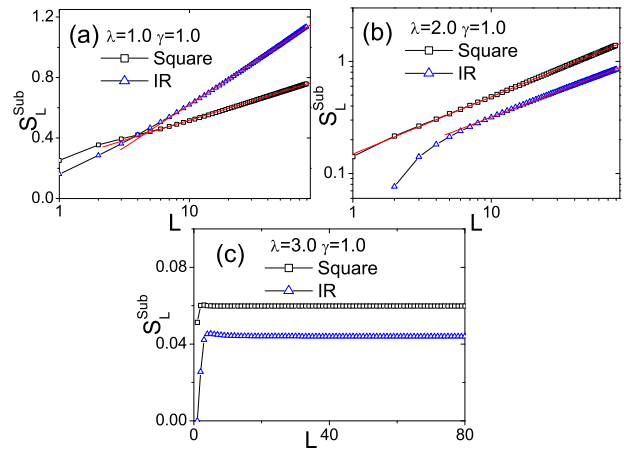}  
\caption{Comparison of $S^{Sub}_L$ for two different partitions. (a)
The phase II : $\lambda=1.0$, $\gamma=1.0$. (b) The boundary regime
: $\lambda=2.0$, $\gamma=1.0$ }
 \label{f4}
\end{figure}

The shape dependence of the logarithmic correction is another way to
distinguish the origin of the non-universal corrections in phase II
from the universal correction at the conformal quantum critical
points\cite{fradkin}.  Here we compare results between square
partitions and isosceles righttriangle (IR) partitions, the latter
is taken as half of a square partition in our numerical simulation.
The results are presented in Fig.\ref{f4}. We find that the choice
of shapes does not change the scaling behavior of $S^{Sub}_L$, i.e.
for phase II the subleading term is still logarithmic divergent
(Fig.\ref{f4}(a)) despite that the prefactor of this logarithmic
term increases significantly in the IR case, and in the boundary
region between phases II and III $S^{Sub}_L$ still follows power-law
scaling behavior (Fig.\ref{f4}(b)). Currently, it is not clear
whether under the change of geometry shape the power factor will
change or not because the second-leading subarea law of $S_L$ is not
known and the inclusion of $\log\log L$ or $\log L$ type terms can
make the fits change significantly.

\begin{table} \caption{\label{tab1} Coefficients of the $S_A$
 : $S_A=a{A}-b{lnA}+const$ in phase II (a) $\lambda=1.0$, $\gamma=1.0$.
 (b) $\lambda=1.0$, $\gamma=2.0$.  }
\begin{center}
\begin{tabular}{cccc}
\hline \hline
Shape & $a$ & $b$ & $const$ \\
\hline
Square (a)  & 0.224170(4) & 0.1189(4) & -0.191(1) \\
Isosceles Righttriangle (a) & 0.253410(6) & 0.2431(8) & -0.003(3) \\
Square (b)  & 0.26230(5) & 0.120(3)&-0.33(1) \\
Isosceles Righttriangle (b) & 0.28719(3) & 0.199(3) & -0.201(9)\\
\hline \hline
\end{tabular}
\end{center}
\end{table}

In fact,  one can fit our data using $S_L\equiv$,
$S_A=a{A}-b{lnA}+c$ to obtain the coefficient directly for phase II.
It is noted that the boundary length $A$ for a square partition is
$4L$ and for an IR partition, $A=(2+\sqrt{2})L$. For both of these
two cases studied in Table I, it is shown that all coefficients are
changing with the partition shape, which suggests that there does
not exist simple relations for different shapes unlike the case in
the 2D $z=2$ conformal quantum critical points\cite{fradkin}.
Interestingly, for the case ($\lambda=1.0$, $\gamma=1.0$), we find
that the prefactor b is equal to 0.1189(4) if the partition is a
square and the prefactor b is equal to 0.2431(8) if the partition is
an IR. This result is very similar to the prediction in
Ref.\cite{fradkin}, as the subarea law
$b^{square}\sim\frac{c}{9}\sim0.111c$ and
$b^{IR}\sim\frac{61c}{252}\sim0.242c$ if we consider $c=1$ here.
However, this is only a special case since one can immediately see
there exist large differences in the ratio of
$\frac{b^{square}}{b^{IR}}$ in other cases, for example,
$\lambda=1.0$, $\gamma=2.0$ shown in Table I.

In summary, our results suggest that there exist significant
differences in the leading subarea term of the EE between
non-critical models and critical models with point nodes in 2D
bipartite fermionic systems. In the non-critical phases the leading
order correction is a non-universal negative constant, proportional
to the square root of correlation length. And in the critical phases
with point nodes we find a subdominant logarithmic additive term
with a non-universal coefficient. In the boundary region between the
point-node and the completely gapped phases the subarea law can not
be described by a simple logarithmic behavior anymore. Our analysis
suggests that there is an unusual non-additive power law relation,
which also depends on the shape of geometrical partition.

Useful discussions with  J. E. Moore,W.Li, T. Roscilde are
gratefully acknowledged. L.T. acknowledge hospitality at Theoretical
division T-11, LANL. Computational facilities have been generously
provided by the HPCC-USC Center. This work was supported by the
Petroleum Research Foundation, grant ACS PRF$\#$ 41757.





\begin{thebibliography}{99}
\bibitem{note01} It has recently been shown that the shape of particular block partition
is reflected in additive contants for the case of topological
phases. In fact, we have observed indications of a similar
phenomenon for the bilinear model studied here. For example the
ratio of the constant term for a 8-corner partition vs. a 4-corner
partition at the case $\lambda=3.0$ and $\gamma=1.0$ in phase III is
found to be exactly 2, as predicted in Ref. \cite{topo04}.
\end{thebibliography}
\end{document}